\documentstyle[12pt]{article}
\newcommand{\be}{\begin{equation}}
\newcommand{\ee}{\end{equation}}
\def\bq{\begin{eqnarray}}
\def\eq{\end{eqnarray}}
\def\n{\nonumber}
\catcode `\@=11
\catcode `\@=12

\title{\bf\huge On product spacetime with 2-sphere of constant curvature}
\author{Naresh Dadhich\thanks{E-mail : 
nkd@iucaa.ernet.in} \\
{\sl Inter-University Centre for Astronomy \& Astrophysics,}\\
{\sl Post Bag 4, Ganeshkhind, Pune - 411 007, India.}} 
    \date{}
    \input epsf
    \begin{document}
    \maketitle
    
     \begin{abstract} 
 If we consider the spacetime manifold as product of a constant curvature 
2-sphere (hypersphere) and a 2-space, then solution of the Einstein equation 
requires that the latter must also be of constant curvature. There exist 
only two solutions for classical matter distribution which are given by the 
Nariai (anti) metric describing an Einstein space 
and the Bertotti - Robinson (anti) metric describing a uniform electric 
field. These two solutions are transformable into each other by letting 
the timelike convergence density change sign. The hyperspherical solution is 
anti of the spherical one and the vice -versa. For non classical matter, we 
however find a new solution, which is electrograv dual to the flat space, and  
describes a cloud of string dust of uniform energy density. We also discuss 
some interesting features of the particle motion in the Bertotti - Robinson 
metric.
    \end{abstract}

    \n PACS numbers : 04.20,04.60,98.80Hw

    \newpage 

 We consider the 4-dimensional spacetime as a product manifold of two 
spaces  $R^2 \times S^2$ with $S^2$ having constant curvature. The latter condition 
implies that  $R^2_2 = R^3_3 = const.$ and $R^0_0 = R^1_1$ where the 
coordinates are designated as ${t=0, z=1, \theta=2, \varphi=3}$. Each 
space is a 2-space intersecting the other orthogonally and hence each 
would be specified by a single curvature, which would be $R^{23}~_{23}$ 
for $S^2$ and $R^{01}~_{01}$ for $R^2$. The Einstein equation would hence 
be a relation between them and it turns out that the only classical 
matter distribution it can sustain is the Einstein space ($R_{ab} = 
\Lambda g_{ab}, \rho + p = 0$) or the uniform electric field, with $R = 0$. 

 Let us define the the energy density by $\rho = T_{ab}u^au^b$, timelike 
convergence density by $\rho_t = (T_{ab} - 1/2Tg_{ab})u^au^b$ and the null 
convergence density by $\rho_n = T_{ab}k^ak^b$ where $u^au_a = 1, k^ak_a = 
0$ [1]. Note that for the product manifold, $\rho = R^2_2, \rho_n = R^0_0 
- R^1_1 = 0, \rho_t = - R^0_0$. For the classical matter distribution, 
there exist the only two possibilities; (i) $\rho + \rho_t = 0$, the 
Einstein space given by the Nariai metric [2] and (ii) $\rho - \rho_t = 0$, 
the uniform electric field given by the Bertotti - Robinson metric [3,4], 
and $\rho_n = 0$ always. The only other possibility could be of the two 
curvatures one could vanish; i.e. either $\rho = 0$ or $\rho_t = 0$. The 
former cannot vanish because it defines the constant 
curvature of the 2-sphere (hypersphere), the construction we began with. 
Even for the non classical matter, the only possibility is $\rho_t = 0$, 
which is the equation of state for the cosmological defects, string dust, 
global monopole and global texture [5]. The new solution that we shall 
obtain could describe a string dust of constant energy density. The 
solution could be shown to be the electrograv dual to the flat space [6,7].

 For defining the electrogravity duality, we decompose, in analogy with 
the Maxwell theory, the Riemann 
curvature into electric and magnetic parts relative to a timelike unit 
vector. Since the Riemann curvature is a double 2-form, and hence it 
would always be the double projection. The active electric part is given 
by $E_{ab} = R_{acbd}u^cu^d$, the passive electric part by $\tilde E_{ab} = 
*R*{acbd}u^cu^d$ and the magnetic part by $H_{ab} = *R_{acbd}u^cu^d$ 
where $*R*_{abcd} = 1/4\eta_{abmn}\eta_{cdpq}R^{mnpq}$ and $\eta_{abcd}$ is 
the 4-volume element. In terms of the electromagnetic parts, the 
Ricci curvature is given by
\be
R_{ab} = E_{ab} + \tilde E_{ab} + (E + \tilde E)u_au_b - \tilde Eg_{ab} + 
u^cH^{mn}(\eta_{acmn}u_b + \eta_{bcmn}u_a)
\ee
where $E =E^a_a, \tilde E^a_a = \tilde E$, and $\rho = - \tilde E, 
\rho_t = - E$.

 By the electrogravity duality we mean [6],
\be
E_{ab} \leftrightarrow \tilde E_{ab}, H_{ab} \rightarrow H_{ab}
\ee
That is the interchange of active and passive electric parts. Note that 
under duality, the Ricci and the Einstein tensors interchange , which is 
because contraction of the Riemann is Ricci while its double dual is 
Einstein.

 Clearly under the duality transformation, $\rho \leftrightarrow \rho_t, 
\rho_n \rightarrow \rho_n$ which means the Einstein space would be anti 
dual while the uniform electric field solution would be self dual. We 
would further show that the new solution describing a string dust would 
be dual to the flat space. 

 We write the metric for the product spacetime as
\be
 ds^2 = c^2 dt^2 - a^2 dz^2 - \frac{1}{\lambda^2} (d\theta^2 + sin^{2}\theta  
 d\varphi^2)
\ee
where $\lambda$ is a constant with dimension of inverse length, and $c, a$ 
are in general functions of $z$ and $t$ which run from $-\infty$ to 
$\infty$. There 
are only two independent components of the Ricci tensor which read as
\be
- \rho_t = R^0_0 = R^1_1 =\frac{1}{c^2}(\frac{\ddot a}{a} - 
\frac{\dot a\dot c}{ac}) - \frac{1}{a^2}(\frac{c^{\prime\prime}}{c} - 
\frac{a^{\prime}c^{\prime}}{ac}), \hspace{0.3cm}
\rho = R^2_2 = R^3_3 = \lambda^2 
\ee

 The two classical matter solutions would be obtained by solving the 
single equation $\rho = \lambda^2 = \pm\rho_t$. For its solution we will 
have to choose the one of $a$ and $c$ or a relation between them. There 
could exist both static and non static solutions but one could always be 
transformed into the other. We shall give the both in the gauge $ac = 1$. 
The solutions are given as follows.

I. $\rho + \rho_t = 0$: The Nariai metric [2] for the Einstein space,
\be
c^2 = a^{-2} = (1 + \lambda^2t^2)^{-1}, ~ 1 - \lambda^2z^2
\ee
which gives $\rho = \lambda^2 = - p$.

II. $\rho = \rho_t = \lambda^2$: The Bertotti - Robinson metric [3,4] for 
the uniform electric field,
\be
c^2 = a^{-2} = (1 - \lambda^2t^2)^{-1}, ~ 1 + \lambda^2z^2
\ee
The electric field would be along the $z$-axis, $F_{01} = \pm \sqrt{2} ac
\lambda$, which would imply
\be
(F^{ij} \sqrt {-g})_{,j} = 0
\ee         
where a coma denotes ordinary derivative. Thus the electric field is 
uniform and is given by $\pm\sqrt{2} \lambda$. This equation indicates 
absence of charges, which would be lying at $z = \pm\infty$ to produce a 
uniform field.  

 Let us note some of the interesting properties of the two solutions:

(a) $N \leftrightarrow BR$ when $\rho_t \leftrightarrow -\rho_t$, where N 
stands for the Nariai metric and BR for the Bertotti - Robinson metric. 
That means changing the sign of $\lambda^2$ only in $c$.

(b) The anti metrics of N and BR would be given by letting sphere 
($sin\theta$) to go to hypersphere ($sinh\theta$) in the metric. That is, 
in N and BR let both $\rho \rightarrow -\rho, \rho_t \rightarrow -\rho_t$ 
to get to their anti counterparts.

(c) The anti metrics are transformable into each-other as in (a) above by 
letting $\rho_t \leftrightarrow -\rho_t$.

(d) Under the duality transformation, $\rho \leftrightarrow \rho_t$, the 
Einstein space is anti dual, and the Nariai and anti Nariai are dual of 
each - other. On the other hand, the trace free matter field is self 
dual, and the BR and anti BR are dual of each other with both $\rho$ 
and $\rho_t$ changing sign. This is because under the duality 
gravitational constant changes sign [8].

 Now let us turn to the only remaining possibility of a curved spacetime 
for the metric (3),e.g. $\rho_t = - R^0_0 = 0$. That is $R^2$ is flat 
which would mean $c = a = 1$. The metric (3) would still be curved giving 
rise to the stresses $\rho = T^0_0 = T^1_1 = \lambda^2$ and the rest being 
zero. The equation of state, $\rho_t = 0$ is characteristic of topological 
defects; string dust, global monopole and global texture [5]. In the 
present case it is the string dust [9] that has constant energy density. The 
anti string dust metric would result if we let $sin\theta \rightarrow 
sinh\theta, sphere \rightarrow hypersphere$ in the metric (3) with $a = c= 
1$. This will have $\rho = - \lambda^2$.

 The string dust solution can be obtained from the equation,
\be
E_{ab} = 0, ~\tilde E^a_b = - \lambda^2 g^a_1g^1_b  
\ee
which is dual to the effective flat space equation,
\be
\tilde E_{ab} = 0, ~E^a_b = - \lambda^2 g^a_1g^1_b
\ee
for the metric (3). For a general spherically symmetric metric (replace 
$z$ by $r$, $1/\lambda$ by $b$ and take $a,b,c$ as functions of $r$ and 
$t$) it can be shown after considerable manipulation [10], that the above 
equation characterizes flat space. This shows that the string dust which 
is the solution of the dual set (8) is dual to flat space which is 
the solution of the set (9).

 The stress tensor for string dust is given by $T^{ab} = 
\epsilon \sigma ^{ac} \sigma^b_c/{-\gamma}^{1/2}$ where $\epsilon$ is the 
proper energy density of the string cloud, $\gamma_{ab}$ is the 
2-dimensional metric on the string world sheet, $\sigma^{ab}$ is the 
bivector associated with the world sheet: $\sigma^{ab} = \epsilon^{AB} 
\partial x^a/\partial\lambda^A\partial x^b/\partial\lambda^B$. Here 
$\epsilon^{AB}$ is the 2-D Levi-Civita tensor ($\epsilon^{01} = - 
\epsilon^{10} = 1$) and $\lambda^A = (\lambda^0, \lambda^1)$, where 
$\lambda^0$ and $\lambda^1$ are timelike and spacelike parameters on the 
string world sheet. In Ref.[9] the stress tensor for a spherically 
symmetric metric has been computed and shown that it has $T^0_0 = T^1_1 
\propto 1/r^2$ which in our case would be proportional to a constant 
because the 2-sphere (hypersphere) has constant curvature. The  matter 
distribution represented by the solution is indeed a cloud of string dust. 
As in other cases its anti version would have $sin\theta$ being 
replaced by $sinh\theta$, and $\rho \rightarrow - \rho$.
 
 We shall next consider some interesting features of particle motion in 
the Bertotti - Robinson metric as given by the following form, 
\be
ds^2 = (1 + \lambda^2z^2) dt^2 - (1 + \lambda^2z^2)^{-1} dz^2 - 
\frac{1}{\lambda^2} (d\theta^2 + sin^2\theta d\varphi^2)
\ee

 Here the red-shifted proper 
acceleration is given by $- \lambda z$ which would be attractive/repulsive 
for $z > 0 (z< 0)$. The most pertinent motion in this spacetime is the 
$z$-motion 
and it would clearly be simple harmonic about the stable state of rest $z 
= 0$. A particle sitting at $z = 0$, which could be chosen anywhere on 
the axis freely, would remain stay put there for ever. That would mean 
that a particle at rest would always remain at 
rest at any $z$ while one in motion would execute simple harmonic 
oscillation about some appropriate $z = 0$ location. This happens because 
as it 
moves on the positive $z$, electrostatic energy lying behind it keeps on 
building up 
to pull it back, then it turns back and the same happens on the other 
side. This is an interesting simple harmonic oscillator which can be set 
up anywhere and is entirely 
maintained for ever by gravity (Fig. 1). Of 
course the catch is that the total 
energy of the system is infinite. It is though not very realistic, yet 
it is an interesting and novel example of a relativistic 
analogue of a simple classical situation.

 What would happen if we consider motion of a charged particle in this 
spacetime? 
The effective potential for the pertinent motion would then be given by
\be
V = q\lambda z + (1 + \lambda^2z^2)^{1/2}
\ee
where $q$ is the charge per unit mass. Without loss of 
generality, we can take $q \ge 0$, because negative $q$ would only imply 
reflection in $z$. Here 
there would be three cases corresponding to $q^2 < 1$, $q^2 > 1$ and
$q^2 = 1$.

 Case (i): For $q^2 < 1$, the potential would have the minimum at 
$\lambda z = - q/\sqrt{1 - q^2}$  and $V_{min} = \sqrt{1 - q^2}$. Particle 
would have oscillatory motion like the neutral particle with $V_{min}$ 
being lowered and its location shifted on the negative $z$-axis. However 
the potential would not be symmetric but would be wider on the left of 
the minimum (Fig. 1). A particle sitting at the minimum of the potential 
would 
remain so for ever. That is, a charged particle in a uniform electric field 
is being kept at stable state of rest by gravity.  

 Case (ii): For $q^2 > 1$, there will be no minimum and $V = 0$ at 
$\lambda z = - 1/\sqrt{q^2 - 1}$. It increases monotonically from negative 
large values to positive large values (Fig. 1). Negative energy orbits were 
first 
encountered in the Kerr spacetime [11] and they were confined to ergosphere 
with its outer boundary at $r = 2M$. By bringing in electromagnetic 
interaction, the effective ergosphere could be extended but its extent 
would always remain bounded [12]. In here we have a situation where 
occurrence of negative energy extends upto infinity in the negative $z$ 
direction. Particles with both positive and negative energy can have a 
bounce and go back to infinity. This unphysical feature is due to 
infinite energy of the spacetime.

 Case (iii): For $q^2 = 1$, there again occurs no minimum and it 
asymptotically tends to zero for negative $z$ (Fig. 1).

 With the constant curvature sphere (hypersphere) as one part of the 
product space, there are only three possible solutions. The two are the 
known solutions; Nariai cosmological metric for the Einstein space 
representing a fluid with $\rho + p = 0$ and the Bertotti - Robinson 
metric for uniform electric field. The third one is new and represents 
non classical matter of string dust of constant energy density. The 
Nariai metric has six parameter symmetry group and is not conformally  
flat. It is shown to be created by quantum polarization of vacuum 
and asymptotically it decays into the de Sitter and the Kasner spacetimes 
[13,14]. The Bertotti - Robinson metric is conformally flat and  has also 
been used to calculate one-loop 
quantum gravitational corrections induced by fields of large mass [15]. 
It would be worthwhile to carry out such calculations for the new 
string dust solution, which could in a way be viewed as "minimally" 
curved because it is completely free of the Newtonian gravity [7].

 Apart from the new solution, the novel feature of the paper is to 
expose the inter - transformability of the Nariai and the Bertotti - 
Robinson metrics, the role of electrogravity duality in finding the new 
solution and some interesting features of particle motion in the 
spacetime of uniform electric field. 
It is interesting that a particle at rest in this spacetime would remain at 
rest for ever while the one in motion along the 
axis of the field would execute simple harmonic oscillation for ever.
 
{\bf Acknowledgement:} I wish to thank Varun Sahni and Ramseh 
Tikekar for enlightening and useful  discussions, and Ivor 
Robinson for reading the manuscript.

\begin{figure}
\centering
{\epsfysize=8.5cm
{\epsfbox[0 0 550 350]{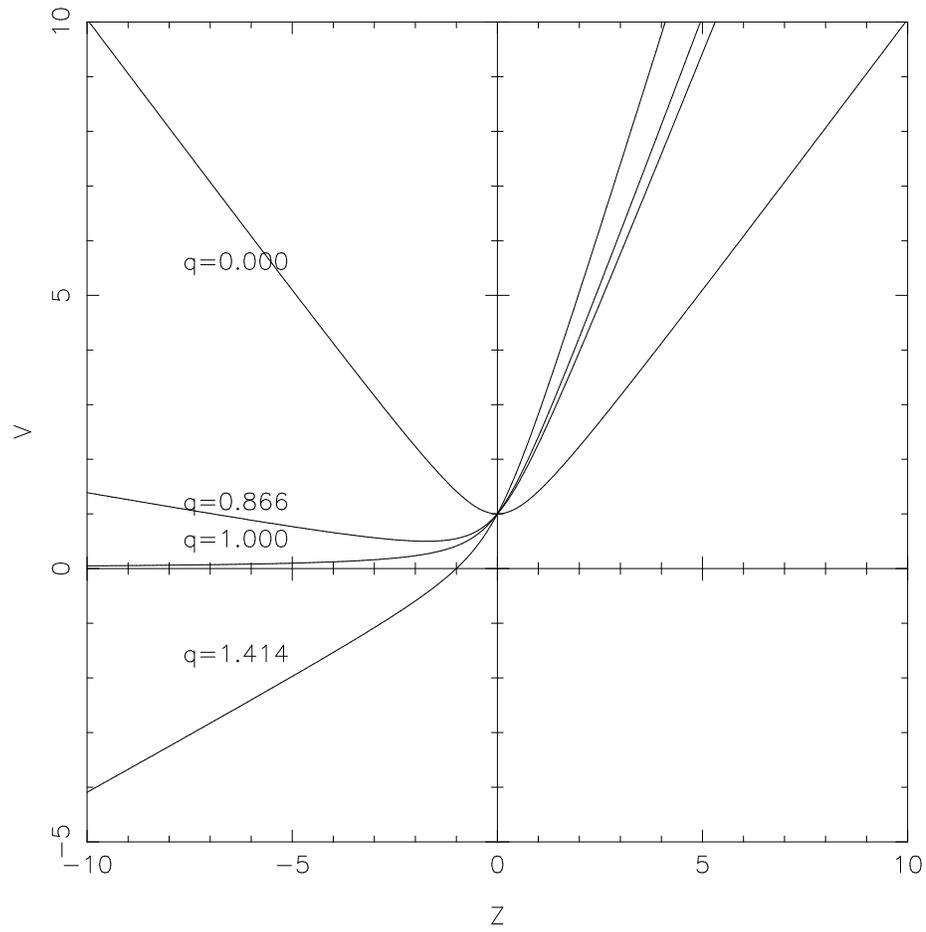}}} 
\caption[]{ Potential plots for various values of $q$}
\end{figure}
   
\end{document}